\documentstyle[aps,psfig]{revtex}
\begin{document}
\draft
\preprint{MKPH-T-98-2}

\title{Analysis of Eta Production using a Generalized Lee Model}
\author{Johannes Denschlag$^{\mbox{$\ast$,\ddag}}$,
 Lothar Tiator$^{\mbox{$\ast$}}$,
 Dieter Drechsel$^{\mbox{$\ast$}}$}
\address{$\ast$ Institut f\"ur Kernphysik\\
Johannes Gutenberg Universit\"at Mainz\\ J. J. Becherweg 45, 55099
Mainz, Germany }
\address{
\ddag ~Institut f\"ur Experimentalphysik\\
Universit\"at Innsbruck\\
Technikerstrasse 25, 6020 Innsbruck, Austria
}
\date{\today}
\maketitle
\begin{abstract}
  We have investigated the processes N($\pi$, $\pi$)N and N($\pi$,
  $\eta$)N close to eta threshold using a simple, nonrelativistic Lee
  model which has the advantage of being analytically solvable. It is
  then possible to study the Riemann sheets of the S-matrix and the
  behavior of its resonance poles especially close to threshold. A
  theoretical simulation of the experimental cusp effect at eta
  threshold leads to a characteristic distribution of poles on the
  Riemann sheets. We find a pole located in the $4^{th}$ Riemann
  sheet that up to now has not been discussed. It belongs to the cusp
  peak at eta threshold. In addition we obtain the surprising result
  using the Lee model that the resonance $S_{11}(1535)$ does not play
  a large role. The main features of the experimental data can be
  reproduced without explicitly introducing this resonance.
  Furthermore, we have also studied the reactions N($\gamma$, $\pi$)N
  and N($\gamma$, $\eta$)N and find reasonable agreement between the
  data and both models with and without the $S_{11}(1535)$ resonance.

\pacs{PACS numbers: 13.60.Le, 14.20.Gk, 13.75.Gx \\
{\em Keywords}: Lee model, $S_{11}(1535)$, eta production}

\end{abstract}

\section{ Introduction }

The advent of new electron accelerators and intense photon sources
substantially improved the data basis of electron and photoproduction
of eta mesons. High quality data for angular distributions and total
cross sections for photon energies between threshold and 790 MeV may
be considered a qualitative breakthrough in the experimental field.
The progress in experimental proficiency demands a better theoretical
understanding of these processes.

The eta meson is believed to be mainly produced by the resonance
$S_{11}(1535)$ which is located close to eta threshold (1486 MeV).
Considerable work has been done in the case of eta photoproduction
\cite{Ben95}, \cite{Sau95}, \cite{Tia94} with the $S_{11}(1535)$
resonance as the main production channel.

The eta threshold also influences scattering cross sections other than
eta production. This is especially true in elastic pion scattering
where a strong cusp effect is observed due to unitarity. The presence
of this cusp makes it difficult to get information on the resonance
$S_{11}(1535)$, because the cusp and the resonance have similar
signatures when analyzing experimental data. Even the existence of
the resonance $S_{11}(1535)$ seems to be questionable \cite{Hoh92},
\cite{Hoh98}. In a recent coupled channel calculation for eta and
kaon photoproduction, Kaiser et al. \cite{Ka97} have shown that the
$S_{11}(1535)$ can as well be explained as a quasi-bound state of
kaon/$\Sigma$-hyperon.

It is generally accepted that the $S_{11}(1535)$ couples strongly to
the eta while the neighboring resonance $S_{11}(1650)$ for all
practical purposes does not decay into eta mesons. It is not clear
why these two neighboring resonances, having the same quantum numbers,
behave so differently.

We investigated theoretically the combination of a resonance and the
threshold effect using a Lee model because of its simplicity and
clearness. It is fully analytically solvable, unitary and can in
principle be extended to an arbitrary number of different mesons and
resonances. The absence of antiparticles makes the model
nonrelativistic. However, a covariant formalism of resonance
excitation also leads to problems, e.g. anti-resonances ($\bar q \bar
q \bar q$) that are very unlikely in terms of quark models.
Analyticity is mainly due to an inherent Tamm-Dancoff approximation in
the model which limits the number of mesons present at any instant.

Our model describes the interaction of the nucleon, the two resonances
$S_{11}(1535)$ and $S_{11}(1650)$, the pion and the eta meson. Making
use of perturbation theory we also included photoproduction of pions
and eta mesons. From our calculation we find that two-pion production
can not entirely be neglected. It was accounted for in a simple
manner as discussed in the sect. \ref{sectphaseanalyse}. Coupling
constants used in our model were determined by fitting to experimental
data from pion scattering experiments and eta (pion) photoproduction.

It is especially interesting to study the scattering matrix S. It is
defined on several Riemann sheets of the complex energy plane. One
finds that poles of S are distributed in a characteristic way on these
sheets and can be identified with objects like resonances or the cusp.

The Lee model was introduced in 1954 by T.D. Lee to study questions
concerning renormalization of field theories \cite{Lee54}. Pauli,
Glaser and K\"all\'en \cite{Gla56}, \cite{Pau55} used this model in order
to investigate the appearance of ``ghost states", states of negative
probability density. Ghost states appear when coupling constants are
chosen to be large. This lead to the discussion of the meaning of an
indefinite metric in a Hilbert space. H\"ohler \cite{Hoh58} used the
Lee model for a quantum mechanical examination of the exponential
decay law of unstable particles. The Lee model has not been
extensively utilized for decades. A good introduction can be found in
\cite{Schwe} and \cite{Lev59}. Using the Lee model our main goal was
not the perfect description of experimental data but to investigate
the interplay and meaning of both cusp effect and resonance
$S_{11}(1535)$ at threshold.

We find that in order to properly treat the threshold effect,
rescattering and calculation to all orders of perturbation theory
is essential. A simple Born approximation at eta threshold is
insufficient. We show which Riemann sheets should be considered
when looking for resonance poles on the complex energy plane. The
sheet structure plays an important role at threshold.

From the fact that the experimental data can be roughly described
without explicitly using the resonance $S_{11}(1535)$, we think that
the importance of this resonance is commonly overestimated and
threshold effects must be taken into account.

In the following we  give an introduction to the generalized Lee model.
We then discuss the fit results together with the experimental data.
First we  consider the production of eta mesons when pions are scattered
by a nucleon. Finally we treat eta and pion photoproduction.

\section{ A Generalized Lee Model }
\label{chapleemodel}

According to the quantum numbers of the $S_{11}(1535)$ resonance, eta
production happens mainly in a s-wave channel. Since angular momentum
and isospin are conserved quantities in strong interaction, we can
also restrict our calculation to the Hilbert subspace where L = 0 and
I = 1/2. Furthermore, we work in the center of momentum system. These
choices considerably simplify our calculations.

The particles that appear in our model are the nucleon N, the two
resonances $R_{1}$ and $R_{2}$ (namely $S_{11}(1535)$ and
$S_{11}(1650)$) and the two mesons $M_{1}$ and $M_{2}$ (pion and eta).
Taking the quantum numbers in isospin space to be $I = 1/2$ and $I_{z}
= 1/2$, the nucleon pion state is actually a superposition of the
physical pion states: 
\begin{equation}
  \mid N, \pi, I=1/2, I_{z}=1/2 \rangle = {1\over \sqrt{3}} \left(
    \sqrt{2} \mid n \rangle \mid \pi^{+} \rangle - \mid p \rangle \mid
    \pi^{o} \rangle \right).
\end{equation} 
We adopt the following notation, so that the isospin does not appear
explicitly:
\begin{equation}
  \mid N, \pi \rangle = \mid N, \pi, I = 1/2, I_{z} = 1/2 \rangle \,.
\end{equation} 
Because we want to make full use of the isospin formalism we have the
difficulty of attributing masses to the pion and the nucleon since the
physical particles ($\pi^{o}$, $\pi^{+}$) and ($n$, $p$) all have
different masses. We chose average masses $m_{\pi}=137$ MeV and
$m_{N}=939$ MeV.

The mesons and the nucleon are treated as stable particles. However,
the resonances $S_{11}(1535)$ and $S_{11}(1650)$ are not stable,
because of the decay into nucleon and mesons. This leads to a
``dressing'' of resonances, e.g. the physical resonances are
surrounded by a cloud of virtual mesons. In our calculation we use the
bare resonance states - not the physical ones. Therefore the masses of
the resonances, $m_{Rj}$, that appear in the model are bare masses and
are used as fit parameters in the model.

\begin{figure}[htb]
\begin{center}
\mbox{\input epsf
 \epsfxsize4.10in \epsfbox{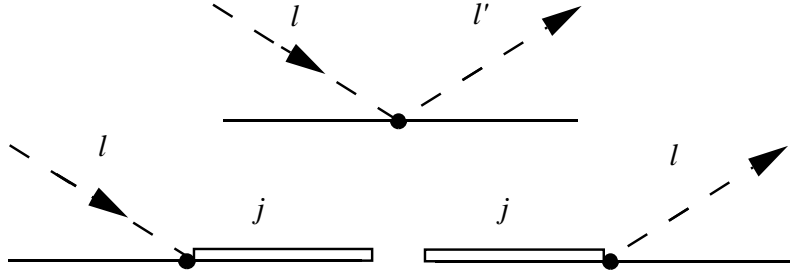}}
 \end{center}
\caption{
  The vertices for the resonance and the contact interaction.  The
  indices $l$ and $l'$ stand for the meson type, the index $j$ refers
  to the resonances.  }
\label{bildvertices}
\end{figure}

The following self adjoint Hamiltonian describes the five particles
and their interactions. It consists of two parts, the free Hamiltonian
and the interaction term:
\begin{eqnarray} 
H = H_{o} + H_{int}\, ,\nonumber
\end{eqnarray}
so that
\begin{eqnarray}
  H_{o} \mid R_{j} \rangle & =& m_{Rj} \mid R_{j} \rangle \,,\nonumber \\ 
  H_{o} \mid N (- \vec k), M_{l} (\vec k) \rangle & =& W_{l} \mid N (-
  \vec k), M_{l} (\vec k) \rangle \,,
\end{eqnarray}
where $W_{l}$ is the total energy in the center of momentum (c.m.)
system, 
\begin{equation} 
W_{l} = \sqrt{ m_{l} ^{2} + \vec k^{2} } + \sqrt{ m_{N}
  ^{2} + \vec k^{2} } .  
\end{equation} 
$\vec k$ is the momentum of particles in the c.m. system. We
distinguish two sorts of interaction among the particles: ``resonance
interaction" and ``contact interaction" (see Fig.
\ref{bildvertices}). In the resonance interaction a meson and a
nucleon combine to form a resonance, or a resonance disintegrates into
a meson and a nucleon
\begin{displaymath}
R_{j} \quad \Longleftrightarrow \quad N, M_{l}\,,
\end{displaymath}
\begin{eqnarray} 
\label{matrixele1}
\langle R_{j} \mid H_{int} \mid N (- \vec k ), M_{l} (\vec k) \rangle
&=& g^{R}_{j,l} ~ F_{l}(W_{l})\,.
\end{eqnarray}
The index $l$ is used for mesons, index $j$ refers to resonances and
$g^{R}_{j,l} $ is the respective coupling constant. $ F_{l}(W_{l}) $
is a real cut-off form factor which is introduced to assure
convergence of expressions that will appear later in the calculations
for the cross sections. It determines how the interaction between the
particles weakens as the kinetic energy rises.

The contact interaction on the other hand corresponds to a
nonresonant interaction, ingoing mesons are scattered into outgoing
mesons,
\begin{displaymath}
N,M_{l} \quad \Longleftrightarrow \quad N, M_{l'}\,,
\end{displaymath}
\begin{eqnarray} \label{matrixele2}
  \langle N, M_{l} ( \vec k) \mid H_{int} \mid N, M_{l'}(\vec k')
  \rangle = { g_{l',l} \over m_{\pi} } ~ F_{l'}(W'_{l'})~ F_{
    l}(W_{l})\,.
\end{eqnarray}
The coupling constants $g_{l,l'}$ for the contact interaction term are
symmetric, e.g. $ g_{l,l'} = g_{l',l}$.

Fig. \ref{bildvertices} shows the vertices for these interactions.
The cut-off function $F_{l}(W_{l})$ is taken to be the same for the
resonance and contact interaction terms. This is arbitrary, but it
simplifies the calculations. In our computer simulations (see sect.
\ref{chapterformfac}) we chose $F_{l}(W_{l})$ to be a simple
analytical function in order to keep computation time short. In
principal, $F_{l}$ can also be deduced from more advanced models by
evaluating the matrix elements in Eqs. \ref{matrixele1} and
\ref{matrixele2}.

To determine the energy eigenstates of the total Hamiltonian $H$, we
write
\begin{eqnarray}
 H ~\mid N, M_{i}(\vec k) \rangle_{+} =
W~\mid N, M_{i}(\vec k) \rangle_{+} \,,
\end{eqnarray}
where $\mid N, M_{i}(\vec k) \rangle_{+} $ is a scattering state which
satisfies the Lippmann-Schwinger equation for ingoing boundary
conditions,
\begin{eqnarray}
  \mid N,M_{i}(\vec k) \rangle_{+} = \mid N,M_{i}(\vec k) \rangle -
  \frac{1}{H_{o}-W- i\epsilon} H_{int}\mid N,M_{i}(\vec k)
  \rangle_{+} ~.
\end{eqnarray}
$M_{i}$ stands for the incident meson ($i=$incident) that at time $t =
- \infty$ comes in as a plane wave. The Lippmann-Schwinger equation
can be solved using the following ansatz,
\begin{eqnarray}\label{glentwicklung}
  \mid N,M_{i}(\vec k) \rangle_{+} = \sum_{j}\beta^{i}_{j} \mid R_{j}
  \rangle +\sum_{l} \int d^{3} k' ~ \alpha^{i}_{l} (\vec k') \mid
  N,M_{l}(\vec k') \rangle\,.
\end{eqnarray}

The coefficients $\alpha$ and $\beta$ in Eq. (\ref{glentwicklung}) can
be obtained by multiplying from the left with $\langle R_{j}\mid,~
\langle N,M_{l}\mid$ and using the orthonormality of the basis. If we
define
\begin{eqnarray}\label{glt}
t^{i}_{l}& =& \int d^{3}k ~ F_{l}(W_{l}) \alpha^{i}_{l}( \vec k)\,,
\end{eqnarray}
we get
\begin{eqnarray}\label{glbeta}
  \beta^{i}_{j} = - \frac{1}{m_{Rj} - W} \cdot
  \sum_{l}g^{R}_{j,l}~t^{i}_{l}\,,
\end{eqnarray}
\begin{eqnarray}\label{glalpha}
  \alpha^{i}_{l}(\vec k) &=& \delta (\vec k - \vec k_{i}) \delta_{l,i}
  + \frac{F_{l}(W_{l})}{W - W_{l} + i\epsilon} \cdot
  \sum_{l'}t^{i}_{l'}G_{l'l} \,,
\end{eqnarray}
where
\begin{eqnarray}
\label{eqcoup}
G_{l'l}&=& \sum_{j}\frac{g^{R}_{j,l'} \, g^{R}_{j,l}} {m_{Rj}- W} -
{ g_{l',l} \over m_{\pi} }\,,
\end{eqnarray}
$G_{l'l}$ can be considered as an effective coupling constant. It is
energy dependent and symmetric, $G_{l'l} = G_{l'l}(W)= G_{ll'}$.

After inserting (\ref{glalpha}) into (\ref{glt}) one obtains:
\begin{eqnarray}\label{gltl}
  \sum_{ l'} h_{ ll'} t^{i}_{ l'} = \delta_{i,l} F_{i}(W)
\end{eqnarray}
with
\begin{eqnarray}\label{hll}
    h_{ ll'}   = \delta_{l,l'} - H_{ l} G_{ ll'}
\end{eqnarray}
and
\begin{eqnarray}\label{gldhl}
  H_{l} = H_{l} (W) = \int d^{3}k ~ \frac{F_{l}^{2}(W_{l})}
  {W_{l} - W - i\epsilon} \,.
\end{eqnarray}
Eq. (\ref{gltl}) is a system of linear equations for $t^{i}_{l}$ which
can be solved by standard mathematical methods. We note that in the
derivation of Eq. (\ref{gltl}) the number of different resonances and
mesons in the model is arbitrary. If for example n different mesons
are involved in the scattering process, a $n \times n$ matrix needs to
be inverted.  When solving for $t^{i}_{l}$ one encounters an important
function which we call $h_{det}$. It is the determinant of $h_{ll'}$
in Eq. (\ref{gltl}) and contains most of the information of the
scattering process. It is a function of $G_{ll'}$ and $H_{l}$. As
will be shown $h_{det}$ as well as $H_{l}$ are meromorphic functions
of $W$. They are defined on several Riemann sheets.

Once the $t_{l}^{i}$ are determined the scattering amplitude $T_{fi}$
can be calculated,
\begin{eqnarray}\label{tfi}
  T_{fi} = ~ \langle N,M_{f}(k_{f}) \mid H_{int} \mid N,M_{i}(k_{i})
  \rangle_{+} = - F_{f}(W) \sum_{ l'} G_{ f l'} ~ t^{i}_{ l'}\,,
\end{eqnarray}
where $f$ and $i$ stand for outgoing and incoming mesons respectively.

\subsection{An example} 
\label{chapsimpleexample}

In order to familiarize the reader with the Lee model, we present here
a simple and instructive example that shows the principal features of
the general Lee model. We will calculate scattering amplitudes, cross
sections and the scattering matrix, which will all appear in their
familiar textbook forms.

We consider the case where the two mesons, $\pi$ and $\eta$, couple
only to one resonance $R_{1}$ (resonance interaction) without contact
interactions. The coefficients $\alpha $ and $\beta $ of the
scattering state are then found to be
\begin{eqnarray}
  \alpha^{i}_{l}(k) &=& \delta (\vec k - \vec k_{i})~ \delta_{l,i}-
  \beta^{i} ~ g^{R}_{1,l}~ F_{l}(W_{l})~ \frac{1}{W_{l} - W
    -i\epsilon},\\ \beta^{i} &=& - g^{R}_{1,i} F_{i}(W )~
  \frac{1}{h_{det}}\,,
\end{eqnarray}
where $h_{det}$ is the determinant mentioned before:
\begin{eqnarray}\label{glha}
  h_{det} = h_{det}(W) &=& m_{R} -W - \sum_{l} ( g^{R}_{1,l} )^{2}
  H_{l}\,.
\end{eqnarray}
We immediately obtain expressions for the scattering amplitude
$T_{fi}$ and the scattering cross section
\begin{eqnarray}
  T_{fi} &=&~ \langle N,M_{f}(k_{f}) \mid H_{int} \mid N,M_{i}(k_{i})
  \rangle_{+}\\ 
  &=&\displaystyle - g^{R}_{1,i} ~ g^{R}_{1,f}
  ~{F_{f}(W) F_{i}(W) \over h_{det}} \,.
\end{eqnarray}
Using Fermi's rule we obtain the cross section
\begin{eqnarray}
  \frac{d\sigma}{d\Omega}&= & \frac{2\pi}{\hbar} \frac{1}{\mid \vec j
    _{in} \mid} \mid T_{fi} \mid ^{2} \varrho (W)\,,
\end{eqnarray}
where $\vec j _{in}$ is the incoming flux and $ \varrho (W) $ is the
phase space density of the outgoing state.

After taking a closer look at the $H_{l}$ we can write the
scattering cross section in a very convenient form. We take $H_{l}$
as in Eq. (\ref{gldhl}) and make use of the identity
\begin{eqnarray}\label{glidentity}
  \frac{1}{W_{l} - W \mp i\epsilon} = P\frac{1}{W_{l} - W} \pm i
  \pi~\delta(W_{l}-W) \,,
\end{eqnarray}
where P stands for the principal value. Let us write
\begin{eqnarray}
  (g^{R}_{1,l})^{2} ~H_{l} &=& \Delta_{l} + i ~ \Gamma_{l}/2 \,,
\end{eqnarray}
where we set
\begin{eqnarray}
  { \Gamma_{l} \over 2} &=& 4 \pi^{2}~ (g^{R}_{1, l})^{2}~
  F_{l}^{2}(W)~k_{l} \frac{\omega_{l} ~ E_{N} }{ W} ~ \Theta(W -
  m_{l} - m_{N})\,, \\ 
  \Delta_{l} &=& 4 \pi ~ (g^{R}_{1,l})^{2} ~ P
  \int_{m_{l}+m_{N}}^{\infty} dW_{l} { F^{2}_{l}(W_{l}) \over W_{l}
    -W } k_{l} \frac{\omega_{l} ~ E_{N} }{ W_{l}}\,,
\end{eqnarray}
and $\omega_{l} = \sqrt{m_{l}^{2} + \vec k ^{2}_{l}} $ and $E_{N} =
\sqrt{m_{N}^{2} + \vec k ^{2}_{l}} $. $\Gamma_{l}$ turns out to be
the partial width of the resonance for the decay channel into meson
$l$ and the nucleon. The step function $\Theta $ in the expression
for $\Gamma_{l}$ leads to a zero partial width below threshold of
meson $l$, and $\Delta_{l}$ can be interpreted as the mass shift of
the resonance due to its interaction with the mesons.

This can be clearly seen when we look at the expression for the
scattering cross section that we finally obtain:
\begin{eqnarray}
  \frac{d\sigma}{d\Omega} &=& \displaystyle {1 \over k_{i}^{2}}
  \left\vert \displaystyle { \displaystyle \sqrt{
        \Gamma_{i}\Gamma_{f}/4} \over h_{det}}\right\vert^{2} \\ &=&
  {1 \over k_{i}^{2}} \displaystyle { \displaystyle
    \Gamma_{i}\Gamma_{f}/4 \over (m_{R}-\Delta_{\pi} - \Delta_{\eta}
    -W)^{2}+ (\Gamma_{\pi}/2+\Gamma_{\eta}/2)^{2}}\,. \nonumber
\end{eqnarray}
The cross section has the well known Breit-Wigner form of a resonance.
We note, that the total Breit-Wigner width of the resonance,
$\Gamma_{tot}$, is the sum of the partial widths,
$\Gamma_{\pi}+\Gamma_{\eta}$. The physical mass of the resonance,
$m_{R}-\Delta_{\pi} - \Delta_{\eta}$, is composed of the bare mass of
the resonance $m_{R}$ and the ``mass shifts" $\Delta_{\pi}$ and
$\Delta_{\eta}$. The physical mass of the resonance is lowered by the
coupling to the mesons. The partial ``mass shifts" of the resonance,
$\Delta_{l}$, add up to give a total mass shift. It is interesting to
note that the same function $F_{l}$ that cuts off the interaction
between the particles also is the cut-off for the width $\Gamma_{l}$.
$F_{l}$ is supposed to be a smooth function that gradually falls off
as $W$ increases. We then reckon that the widths $\Gamma_{l}$ grow
proportionally to $k_{l}(W)$ at their respective $l$-meson threshold
when $W$ is increased. This is in agreement with general scattering
theory that predicts the same behavior for s-wave resonances.

The S-matrix is a $ 2\times 2$ matrix which corresponds to the two
reaction channels.  One finds
\begin{displaymath}
  {\bf S_{o}} = \left( \begin{array}{cc} \displaystyle 1+\frac{i
        \Gamma_{\pi}}{h_{det}} &
      \displaystyle\frac{i\sqrt{\Gamma_{\pi}\Gamma_{\eta}}}{h_{det}}
      \cr
      \displaystyle\frac{i\sqrt{\Gamma_{\pi}\Gamma_{\eta}}}{h_{det}} &
      \displaystyle1+\frac{i\Gamma_{\eta}}{h_{det}} \cr
\end{array} \right).
\end{displaymath}
$S_{o}$ is unitary and the off-diagonal elements are equal which
corresponds to time inversion symmetry. From $S_{o}$ one can easily
determine the scattering phases by comparing it to the textbook form
of a two channel S-matrix:
\begin{displaymath}
  {\bf S_{o}} = \left( \begin{array}{cc} \displaystyle
      \xi~e^{2i\delta_{1}} & \displaystyle
      i\sqrt{1-\xi^{2}}~e^{i(\delta_{1}+\delta_{2})} \cr \displaystyle
      i\sqrt{1-\xi^{2}}~e^{i(\delta_{1}+\delta_{2})} & \displaystyle
      \xi~e^{2i\delta_{2}} \cr
\end{array} \right),
\end{displaymath}
where $\xi$ is the inelasticity and $\delta_{1/2}$ are the elastic
scattering phases.

\subsection{The Form Factor $F_{l}$ } 
\label{chapterformfac}

The cut-off functions $F_{l}$ are constructed so that at low energies
their values are constant. With increasing energy the interaction
between the particles gradually decreases. It seems reasonable that
at the c.m. energy of about two nucleon masses the interaction should
be strongly suppressed.

Although there are many different cut-off functions which fulfill
these features it turns out that in general they produce similar
results.  We selected a cut-off function simple enough so that the
integrals $H_{l}$ could be calculated analytically. Explicitly we
fixed the cut-off function such that
\begin{eqnarray}\label{eqcutofparam}
  H_{l} (W) &=& \int d^{3}k \frac{ F_{l}^{2}(W_{l})}{W_{l} -W
    }\nonumber \\ \label{gldhl2} &=& \sqrt{2 m_{l}} \int_{m_{l}+
    m_{N}}^{C} dW_{l} {\sqrt{W_{l} - m_{l} - m_{N}} \over W_{l} - W
    }\,,
\end{eqnarray}
where $C$ is a cut-off parameter. $C$ was chosen to be 2000 MeV, so
the interaction of the meson with the nucleon stops abruptly for $
W_{l} > C $. We now show that $H_{l}$ (as a complex function) is
defined on two Riemann sheets.

Let us consider $W$ in Eq. (\ref{gldhl2}) as a complex variable.
$H_{l}$ is then two sheeted and has a cut from $m_{l}+m_{N}$ to $C$.
The easiest way to see this is by approaching the real axis from above
and below and making use of the identity (\ref{glidentity}). The
continuation of $H_{l}$ into its second sheet can be done by contour
deformation of the integral and applying Cauchy's residue theorem.
Explicitly we find:
\begin{eqnarray}
  H_{l} (W) \mid_{1. sheet} &=& 2\sqrt{2 m_{l}} \sqrt{C- m_{l} -
    m_{N}} \\ &-& 2 \sqrt{2 m_{l}} \sqrt{m_{l} + m_{N} - W}~ \arctan
  ({\sqrt{C- m_{l} -m_{N}} \over \sqrt{m_{l} + m_{N} - W}})\,,
  \nonumber
\end{eqnarray}
\begin{eqnarray} \label{glarctandiv}
  H_{l} (W) \mid_{2^{nd} sheet} &=& H_{l} (W) \mid_{1^{st} sheet} + ~2
  \pi \sqrt{2 m_{l}} \sqrt{m_{l} + m_{N} -W }\,. \nonumber
\end{eqnarray}

$H_{l}$ diverges at $W = C$ because of the step of the cut-off at $C$.
In our simulations we always stayed well below $C$.

Under the condition $C - m_{l} -m_{N} >> \mid m_{l} + m_{N} - W \mid
$, we can approximate
\begin{eqnarray}
  H_{l} (W) \mid_{2^{nd}sheet} &\approx& 2 \sqrt{2 m_{l}} \sqrt{ C -
    m_{l} -m_{N}} + \pi \sqrt{2 m_{l}} \sqrt{m_{l} + m_{N} - W} \,.
  \nonumber
\end{eqnarray}
The expression $ \sqrt{m_{l} + m_{N} - W} $ is directly related to
the partial width of the resonance $\Gamma_{l}$ as defined in sect.
\ref{chapsimpleexample}.  At threshold, e.g. $ m_{l} + m_{N} = W $,
$\sqrt{m_{l} + m_{N} - W} $ becomes imaginary and grows like the
meson momentum $k_{l}$ with increasing energy $W$.

\subsection{The S-matrix}
\label{chapsmatrix}

\begin{figure}[htb]
\begin{center}
\mbox{\input epsf
 \epsfxsize4.5in \epsfbox{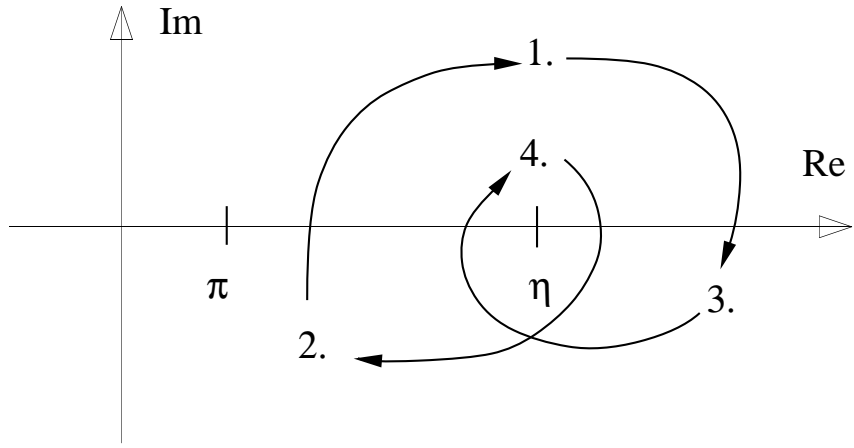}}
 \end{center}
\caption{ \label{bildriemann}
  The four Riemann sheets for the pion and eta meson and the paths
  from one sheet to the next.  }
\end{figure}

\begin{figure}[htb]
\begin{center}
\mbox{\input epsf
 \epsfxsize5.5in \epsfbox{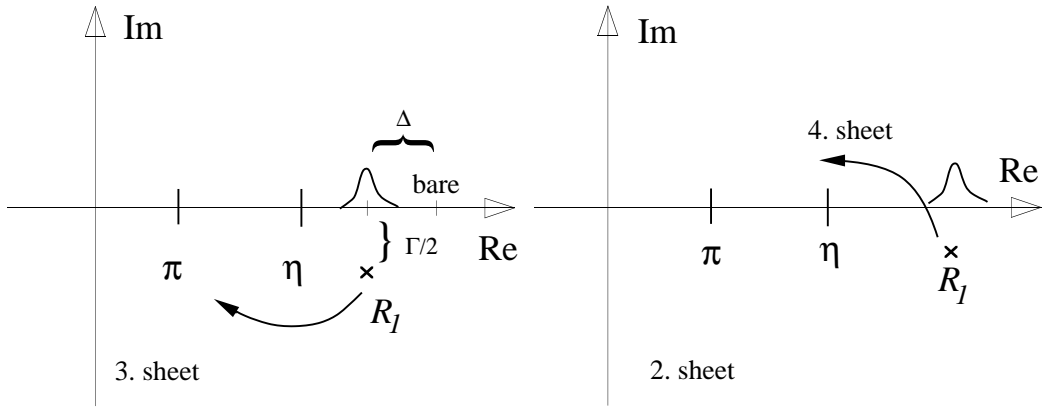}}
 \end{center}
\caption{ \label{bildpolbeweg}
  Pole movement in the $3^{rd}$ and the $2^{nd}$ ($4^{th}$) sheet (see
  text).  }
\end{figure}

We can get a better understanding of the physics underlying the
scattering and production processes by studying the S-matrix $S_{o}$.
We will show in the following how poles of the scattering matrix can
be identified with resonance peaks or cusp peaks in physical cross
sections. It will become clear that the distribution of poles and
their location on different Riemann sheets gives valuable information
on the physics involved. We can take full advantage of the fact that
the Lee model is analytically solvable. Important features of the
S-matrix are determined by general physical laws like unitarity,
causality, time reversal invariance and symmetries of the
interactions. A very good introduction to the subject can be found in
the books by Nussenzveig and Bohm \cite{Nus72}, \cite{Boh93}. In the
following we consider the S-matrix as a complex function of the energy
$W$. S is then found to be meromorphic and is defined on several
Riemann sheets. It is mainly determined by the determinant $h_{det}$,
which is the same for all channels, e.g. in our case $N(\pi,\pi)N$,
$N(\pi,\eta)N$, $N(\eta,\eta)N$ and $N(\eta,\pi)N$.  For our example
in sect. \ref{chapsimpleexample}, we have 
\begin{equation} 
  h_{det} = m_{R}-g^{R}_{1,\pi} H_{\pi} -g^{R}_{1,\eta} H_{\eta}
  -W \,,
\end{equation}
i.e. $h_{det}$ is a four sheeted function, because each $H_{l}$ is
defined on two sheets. There are two cuts located on the real energy
axis corresponding to the two meson thresholds. They start at the
respective meson production thresholds which are also the branch
points for the Riemann sheets. Fig. \ref{bildriemann} shows those four
Riemann sheets of $h_{det}$, the branch points, and the cuts along the
real axis. Whenever a cut is crossed one enters another Riemann sheet.
Starting from the first sheet one moves into the second or third sheet
depending on whether the cut was passed below or above eta threshold.
Passing a sheet twice, brings you back to the original sheet. The
fourth sheet can be reached from the first sheet by passing the cut
twice, once below and once above eta threshold. We find poles in the
complex plane that belong to the resonance, $R_{1}$, called resonance
poles. The coordinates of the S-matrix poles are functions of the
coupling constants. In a typical case when the coupling constants are
small, the resonance poles are found in the second and the third sheet
close to the bare mass $m_{R}$. Following the discussion in sect.
\ref{chapsimpleexample}, the pole coordinates will then be practically
determined by the partial widths $\Gamma_{l}$ and the mass shifts
$\Delta_{l}$. If the resonance peak of the scattering cross section is
located above (below) eta threshold, its shape is determined by the
pole in the {\em third} ({\em second}) sheet. As the coupling
constants are increased the `virtual' meson cloud is bound more and
more tightly to the physical resonance. This lowers the physical mass
of the resonance (the resonance peak moves towards the pion threshold)
and increases its decay width. As a consequence the resonance pole
moves away from the physical axis and towards the pion threshold (see
Fig. \ref{bildpolbeweg}). In general this behavior is the same for
poles in the second or third sheet. However, if the coupling constant
$g^{R}_{1,\pi}$ is held small and constant, and only the coupling of
the eta meson $g^{R}_{1,\eta}$ is increased, the pole in the {\em
  second} sheet moves into the {\em fourth} sheet as drawn in Fig.
\ref{bildpolbeweg}. It is this pole in the fourth sheet that gives
rise to the cusp peak at eta threshold. This will be of importance in
the following section.

We briefly remark that if the coupling of the mesons to the resonance
becomes strong enough the physical resonance finally becomes a bound
state. This bound state is then represented by a pole in the first
sheet on the real axis below  pion threshold. Indeed, with increasing
coupling constants the pole of the second sheet moves towards (and
finally onto) the pion threshold where it can  pass across the branch
point to the first sheet. This behavior of the pole was first seen by
H\"ohler \cite{Hoh58} (see also in detail \cite{Den94} and
\cite{Ma53}).

\section{Scattering phase analysis of pion nucleon scattering}
\label{sectphaseanalyse}

Because we work in the Hilbert subspace corresponding to quantum
numbers $ I=1/2 $, $J = 1/2$ and $L=0$, it is convenient to use a
partial wave phase analysis of the experimental data to determine the
parameters of our model. There is a variety of scattering phase
analysis which differ partially in their results because the
evaluation methods and the underlying data are often not identical.
Among the best known are those from Karlsruhe Helsinki (KH,
\cite{Hoh79}, \cite{Hoh92}, \cite{LaB93}), Carnegie-Mellon-Berkeley
(CMB, \cite{Cut79}, \cite{Cut80}) and from the Virginia Polytechnic
Institute (VPI, \cite{ArnRo}). In our work we used mainly the VPI
analysis by Arndt and Roper which has the convenience of being
supported by the SAID interactive dial-in program.

\begin{figure}[bht]
\begin{center}
\mbox{\input epsf
 \epsfxsize4.0in \epsfbox{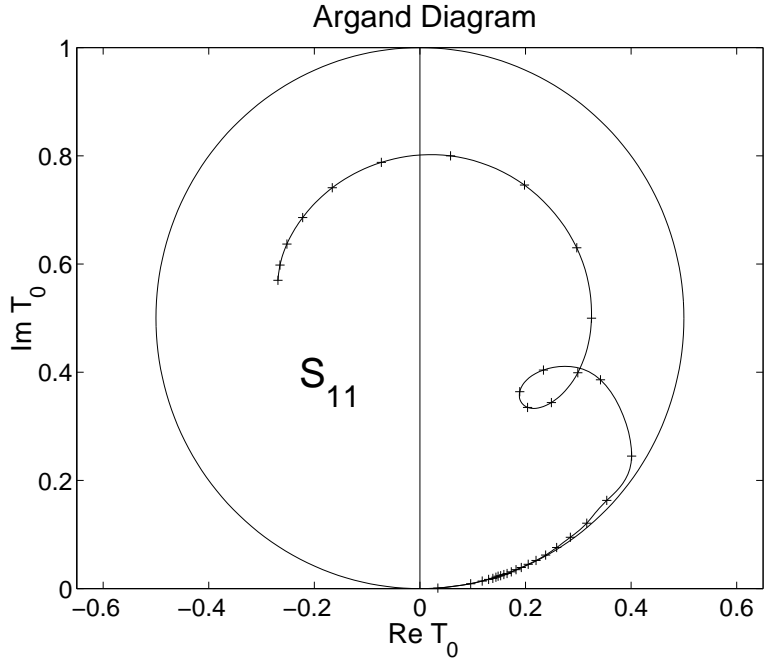}}
 \end{center}
\caption{\label{abbmanleyargand1}
  Argand diagram for the $\pi N$ partial wave $S_{11}$. Data from VPI
  \protect\cite{ArnRo}.  }
\end{figure}

A typical Argand diagram ($T_{0}$) of the $S_{11}$ partial wave for
pion nucleon scattering is depicted in Fig. \ref{abbmanleyargand1}.
The data are taken from a VPI solution (Fall 93) \cite{ArnRo}. Useful
information can be obtained by studying the Argand diagram which shows
how the complex scattering amplitude $T_{0}$ for pion nucleon
scattering changes as the c.m. energy is increased. At low scattering
energies we get $T_{0} = 0$. With increasing energy $T_{0}$ first
moves counterclockwise on the `unitarity' circle which implies purely
elastic scattering of the pion. At about the ``4 o'clock'' position,
$T_{0}$ bends sharply inwards. This corresponds to the threshold of
eta meson production. It is also at this energy that the resonance
$S_{11}(1535)$ is located. The more $T_{0}$ moves into the center of
the unitary circle, the more the scattering is inelastic, in our case
due to the production of eta mesons and two-pion production. After a
small closed loop $T_{0}$ continues in a nearly perfect half circle.
It is those circular patterns in Argand diagrams that indicate the
existence of a resonance \cite{Nus72,Boh93}. In our case it
corresponds to the resonance $S_{11}(1650)$. However, there is no
circle that corresponds to the resonance $S_{11}(1535)$. This may
imply that either there is no resonance or that its circle is heavily
deformed by the opening of the eta meson channel. We address this
question in the next section when we fit two models, one with and the
other without resonance $S_{11}(1535)$ to the experimental values of
$T_{0}$.

\begin{figure}[htb]
\begin{center}
\mbox{\input epsf
 \epsfxsize4.5in \epsfbox{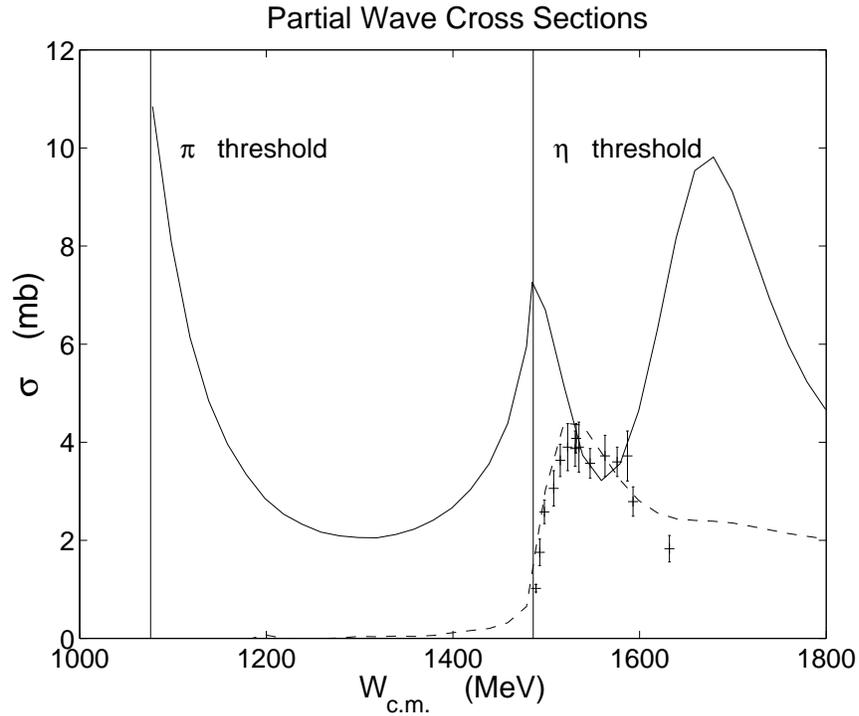}}
 \end{center}
\caption{\label{abbmanleywq}
  Scattering cross sections for the partial wave $S_{11}$, calculated
  from the partial wave analysis by Arndt and Roper
  \protect\cite{ArnRo}.  (-----) elastic cross section $\sigma_{s}$
  for $\pi N \rightarrow \pi N$, (- - - - -) inelastic cross section
  $\sigma_{r}$.  The data points are the cross sections for eta
  production \protect\cite{Nef92}. }
\end{figure}

Fig.\ref{abbmanleywq} shows the calculated elastic ($\sigma_{s}$) and
inelastic ($\sigma_{r}$) partial wave scattering cross sections for
the partial wave $S_{11}$ which have been obtained by
\begin{eqnarray} \label{sigmassigmar}
\sigma_{s} &=& {4 \pi \over k^{2} } \mid T_{0}\mid ^{2} \\
\sigma_{r} &=& { \pi \over k^{2} } \big( 1 - \mid 1 + 2iT_{0}\mid ^{2}
 \big)\,. \nonumber
\end{eqnarray}
The elastic cross section shows two peaks at about 1500 MeV and 1700
MeV. They can be linked to the resonances $S_{11}(1535)$ and
$S_{11}(1650)$ respectively. The peak at 1500 MeV is located at the
eta meson production threshold. Its tip is pointed; it has a cusp.
The formation of the cusp is a threshold effect and is due to
unitarity \cite{Lan79}. We show in the next section that even the
entire corresponding peak at 1500 MeV can be completely explained as a
threshold effect. The peak can be reproduced in a model without the
resonance $S_{11}(1535)$.

In Fig. \ref{abbmanleywq} we show the elastic and inelastic cross
sections together with the data of Clajus and Nefkens \cite{Nef92} for
the total cross section of eta production. From this comparison we
find that the inelasticity of the pion scattering is mainly due to the
production of the eta meson. For higher energies, however, the
multi-pion production is no longer negligible and consists essentially
of two-pion production in the energy range below 2 GeV.

Therefore, in order to reproduce the experimental data qualitatively,
the Lee model should include two-pion production. However, the
inherent Tamm-Dancoff approximation only allows one meson to be
present at a time. We circumvented this by introducing an additional
``meson particle'' in our model representing a two-pion system.

\begin{figure}[htb]
\begin{center}
\mbox{\input epsf
 \epsfxsize3.5in \epsfbox{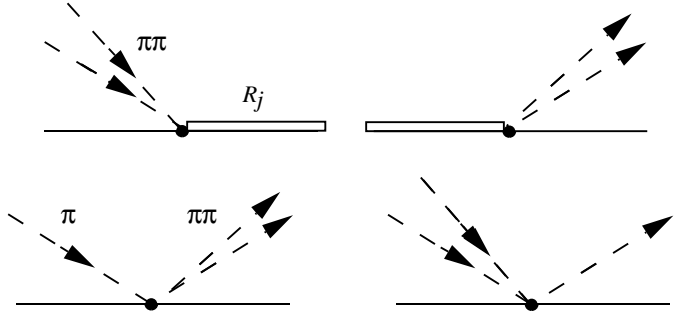}}
 \end{center}
\caption{\label{abbzweipi}
  Vertices for the two-pion system.}
\end{figure}

We then solve the Lee model as before, only now with three different
types of mesons instead of two. The third meson type stands for the
pion pair. The vertices in Fig. \ref{abbzweipi} show the coupling the
two-pion pair to the nucleon and the resonances. This introduces
three additional coupling constants: $g^{R}_{1,2\pi},~
g^{R}_{2,2\pi},~ g_{2\pi,\pi}.$ At the two-pion threshold the phase
space density for the two-pion particle should not increase
proportionally to the momentum $k$ like for the pion or eta meson at
threshold. Instead for a two-pion system the phase space density
grows approximately quadratically with energy:
\begin{eqnarray}
  \mbox{phase space} \propto (W_{2\pi} - 2\mu_{\pi} - m_{N})^{2}\,.
\end{eqnarray}
We can take this into account by choosing $H_{2\pi}$ to be
\begin{eqnarray}\label{gldhpipi}
  H_{2\pi} = \int_{2\mu_{\pi} + m_{N}}^{C_{2\pi}} dW_{2 \pi} {(W_{2
      \pi}-2\mu_{\pi} -m_{N})^{2} \over \mu_{\pi} ~ (W_{2 \pi} -W
    -i\epsilon) }\,.
\end{eqnarray}
The cut-off mass, $C_{2\pi}$, was chosen to be 2000 MeV in our
calculation. With our method of representing two pions by one
effective particle we can not describe exclusive two-pion production,
but rather use this representation to explain the total reaction cross
section.

\begin{figure}[htb]
\begin{center}
\mbox{\input epsf
 \epsfxsize3.5in \epsfbox{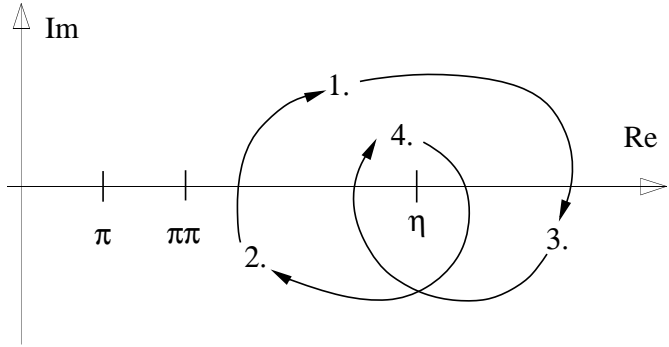}}
 \end{center}
\caption{ \label{abbzweipiblatt}
  Riemann sheets when two-pion production is taken into account.}
\end{figure}

By taking into account two-pion production we also change the analytic
structure of the $S$-matrix. Now there is a third threshold at
$2m_{\pi} + m_{N} $ and additional Riemann sheets appear. Moreover
with our choice of $H_{2\pi}$ in Eq. (\ref{gldhpipi}), $H_{2\pi}$
is no longer a two sheeted function of $W$, but has an infinite
number of sheets. However, these sheets do not play an important role
in our discussion, because we are only interested in the sheet
structure in the immediate vicinity of the eta threshold. The local
sheet structure at eta threshold stays unchanged. Also our discussion
of the Riemann sheets and the motion of poles in sect.
\ref{chapsmatrix} is still valid. Fig. \ref{abbzweipiblatt} shows the
new Riemann sheet structure. The sheets are numbered such that the
sheet structure around the eta threshold looks the same as in Fig.
\ref{bildriemann}.

\section{Determination of coupling constants and poles}
\label{kapbestkopplung}

A glance at the partial wave cross section in Fig. \ref{abbmanleywq}
shows that at pion threshold (far away from the first resonance) the
cross section undergoes a rapid change with increasing energy. The
elastic pion nucleon scattering cross section at pion threshold is
even larger than the resonance peaks of the resonances $S_{11}(1535)$
and $S_{11}(1650)$. The great magnitude of the elastic scattering
cross section is theoretically confirmed by low energy theorems for
pion scattering which make use of chiral symmetry \cite{Tom66},
\cite{Wei66}. Given the nearly constant form factors at threshold as
discussed earlier, it is impossible to simulate the rapid fall-off of
the cross section from pion threshold to an energy of about 1300 MeV
(see Fig. \ref{abbmanleywq}). A better form factor for the
pion-nucleon interaction would have to change quite rapidly at pion
threshold. As a consequence we used our model only at energies above
1350 MeV.

\subsection{Coupling constants}

\begin{figure}[htb]
\begin{center}
\mbox{\input epsf
 \epsfxsize4.5in \epsfbox{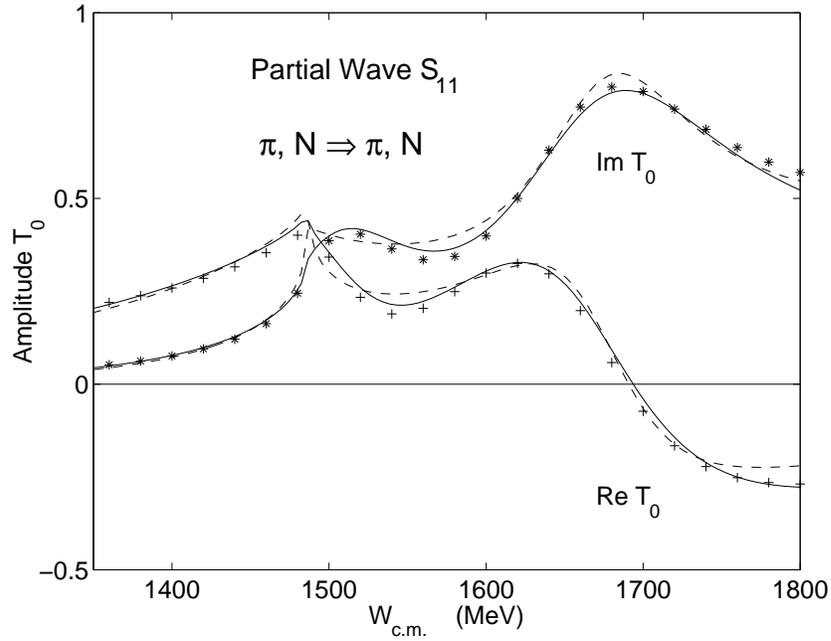}}
 \end{center}
\caption{\label{abbimreresonanz}
  Real and imaginary part of the scattering amplitude $T_{0}$
  ($S_{11}$-partial wave) for the model with $S_{11}(1535)$ (---) and
  without $S_{11}(1535)$ (-~ -~-). }
\end{figure}

\begin{figure}[htb]
\begin{center}
\mbox{\input epsf
 \epsfxsize4.5in \epsfbox{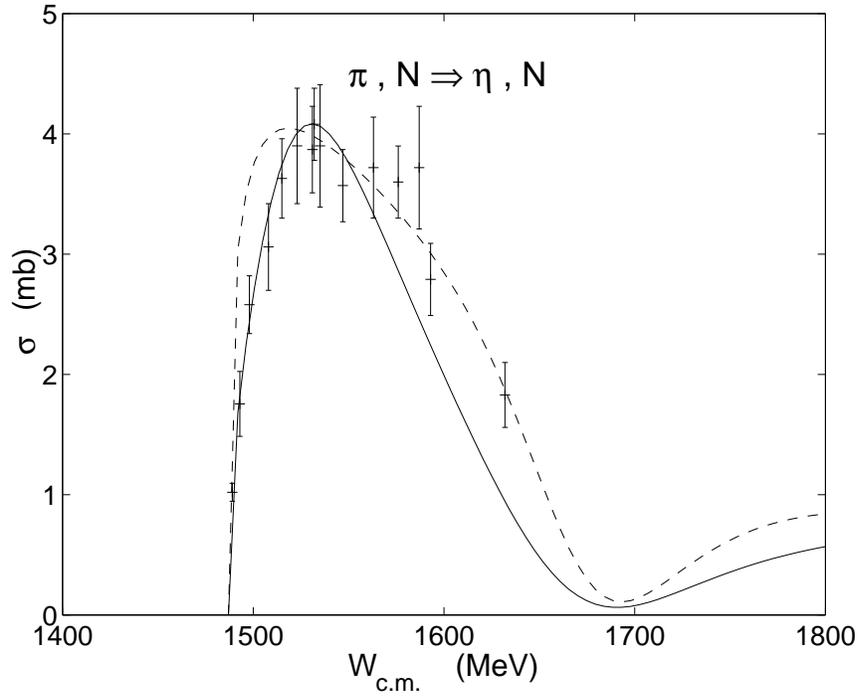}}
 \end{center}
\caption{\label{abbwirku}
  The cross section for eta production in pion nucleon scattering.
  The solid and dashed lines show the model calculations with and
  without resonance $S_{11}$(1535).  Experimental data are taken from
  Nefkens \protect\cite{Nef92}.  }
\end{figure}

The free parameters of the model such as the coupling constants and
the resonance masses were determined by fitting the Lee model to the
experimental data extracted from a partial wave analysis. We fitted
two generalized Lee models, one including the resonance
$S_{11}(1535)$, the other without. After fitting we compared the
fitting results for these two models. From our calculations we observe
that the peak at eta threshold is almost entirely due to the cusp
effect. Even without explicitly including the resonance $S_{11}(1535)$
in the model the peak appears. Moreover we find that the experimental
data we used in our fits can be qualitatively reproduced. The
nonresonant coupling to the eta meson gives similar results as
obtained with a resonance interaction although the model has four fit
parameters less than the model with resonance $S_{11}(1535)$. Fig.
\ref{abbimreresonanz} depicts the complex scattering amplitude $T_{0}$
together with our fitted theoretical curves. The data have been taken
from the VPI solution (SM90) \cite{ArnRo}. A persistent difference
between our two models is found at eta threshold (1486 MeV). Im
$T_{0}$ as obtained with the resonance $S_{11}(1535)$ is much smoother
near threshold than without the resonance. Fig. \ref{abbwirku} shows
the cross section for eta production in pion nucleon scattering.
 
In Tab. \ref{tableconstants} the coupling constants are given for two
fits, Lee model with and without the resonance $S_{11}$(1535).
Because our model is quite different from other standard dynamical
models, the coupling constants of the Lee model can not be directly
compared to the listed standard coupling constants.

Also within the Lee model the magnitudes of resonance coupling
constants cannot be easily compared to the magnitude of contact
couplings. This may be understood when looking at their different
definitions in sect. \ref{chapleemodel}. In order to reproduce the
strong cusp effect and the strong production of eta mesons, the
overall coupling to the eta mesons has to be strong. In one model the
eta meson couples strongly to the resonance $S_{11}(1535)$ (coupling
constant $g^{R}_{1,\eta}$) in the other model it couples strongly by
contact interaction ($g_{\eta,\pi}$). In both models the resonance
$S_{11}(1650)$ couples very weakly to the eta meson while it couples
strongly to the pion.

\subsection{Poles on the Riemann sheets}

\begin{figure}[htb]
\begin{center}
\mbox{\input epsf
 \epsfxsize4.2in \epsfbox{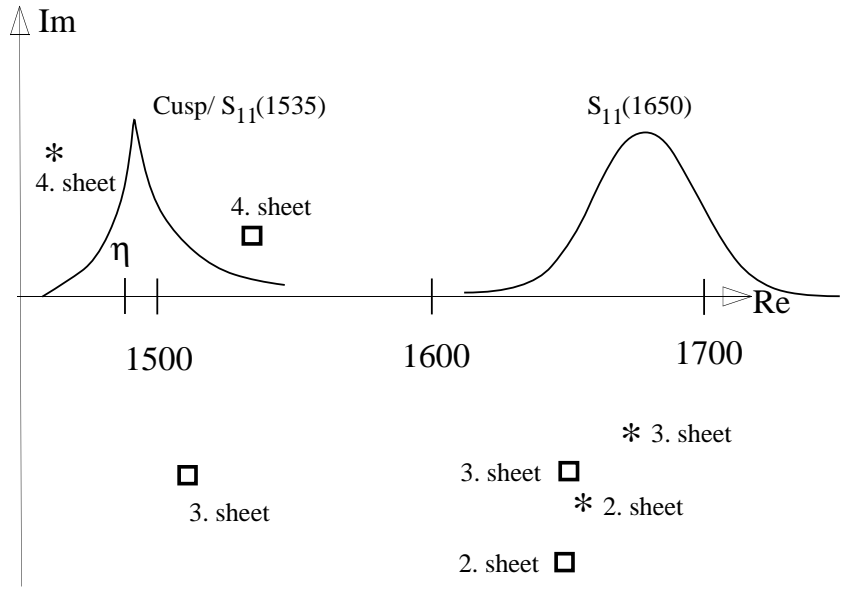}}
 \end{center}
\caption{\label{polbild}
  The pole positions on the Riemann sheets for our two models.
  ($\ast$) model with $S_{11}(1535)$, ($\Box$) model without
  $S_{11}(1535)$.  }
\end{figure}

One can get more information about the analytic structure of the
$T$-matrix and the peaks of the cross section in Fig.
\ref{abbmanleywq} by looking at the pole distribution on the Riemann
sheets. Tab. \ref{tabpolstelle} gives the coordinates of the poles
for the two different models and Fig. \ref{polbild} shows their
locations in the complex plane. We note that the resonance
$S_{11}(1650)$ in both models has poles on the $2^{nd}$ and the
$3^{rd}$ sheets, as is to be expected from sect. \ref{chapsmatrix}.
The coordinates of the poles agree well with the mass and the width of
the resonance $S_{11}(1650)$. In both models we find a pole on the
$4^{th}$ sheet close to the eta threshold which is responsible for the
cusp peak. This pole has not previously been discussed. No additional
resonance pole in the $3^{rd}$ or $2^{nd}$ sheet is needed to account
for the cusp peak.  We can trace back the origin of this pole by
numerically lowering the respective eta meson coupling constants as
described in sect. \ref{chapsmatrix}. The pole then moves back to its
original starting point, which is different for the two models. For
the case of the model with resonance $S_{11}(1535)$ the pole is
originally a resonance pole in the $2^{nd}$ sheet that has been pushed
into the $4^{th}$ sheet by the strong coupling to the eta meson,
whereas in the model without resonance $S_{11}(1535)$ the pole always
stays in the $4^{th}$ sheet.

Another pole in the $3^{rd}$ sheet for the case of the model with
resonance $S_{11}(1535)$ is the resonance pole of the resonance
$S_{11}(1535)$.

Finally we have compared our direct determination of the poles with
the speed plot technique suggested by H\"ohler \cite{Hoh92}. The pole
of the resonance $S_{11}(1650)$ in the $3^{rd}$ sheet is very well
reproduced by the speed technique and found as (1666-81i) MeV in our
model with $S_{11}(1535)$ and as (1671-57i) MeV in our model without
$ S_{11}(1535) $. In addition in both our models we see a very narrow spike with a
width of a few MeV at the $\eta$ threshold which corresponds to the
cusp. The resonance pole of the $S_{11}(1535)$ in our model does not
show up using the speed technique.  It is entirely covered by the cusp
effect.

\section{Eta and pion photoproduction}

\begin{figure}[htb]
\begin{center}
\mbox{\input epsf
 \epsfxsize4.0in \epsfbox{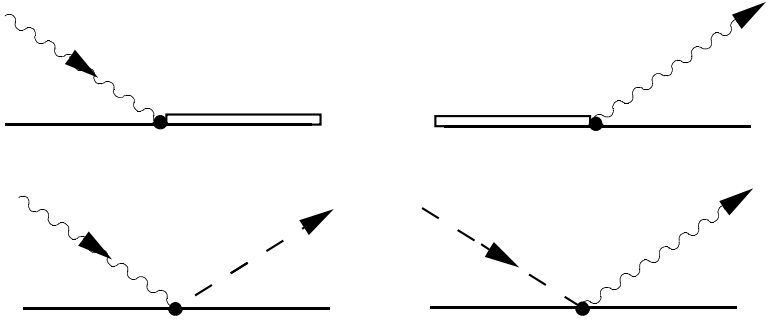}}
 \end{center}
\caption{\label{abbphotovertex}
  Vertices for the coupling of the photons to the nucleon and the
  mesons.}
\end{figure}

In order to check the consistency of our model and to further study
the role of the resonance $S_{11}(1535)$, we studied eta and pion
photoproduction. Since the photon coupling is weak, we use
perturbation theory for the photoproduction. The strong interaction is
still fully accounted for as we can employ the full analytical
solution of the strong interacting particles discussed in the
preceding sections. Fig. \ref{abbphotovertex} shows how the photon
couples to the hadrons via resonance and contact interactions.  The
matrix elements are defined as follows:
\begin{eqnarray}
\langle R_{j} \mid H_{\gamma}
\mid N (- \vec \nu ), \gamma (\vec \nu), \vec \epsilon \rangle
= {g^{R}_{j,\gamma} \over \sqrt{m_{\pi}} }
~\vec \sigma \cdot \vec \epsilon \,,
\end{eqnarray}
\begin{eqnarray}
  \langle N (- \vec \nu), \gamma (\vec \nu),\vec \epsilon \mid
  H_{\gamma} \mid N ( - \vec k), M_{l} (\vec k) \rangle = {
    g_{l,\gamma} \over \sqrt{ m_{\pi}^{3} } }~ F_{l}(W_{l}) ~\vec
  \sigma \cdot \vec \epsilon \,,
\end{eqnarray}
where $j$ is again the index for the resonances, $g^{R}_{j,\gamma}$
and $g_{l,\gamma}$ are the coupling constants for resonance and
contact terms, respectively.  The momentum and the polarization vector
of the photon are denoted $\vec \nu$ and $\vec \epsilon $, and $\vec
\sigma$ is the spin of the nucleon.

\begin{figure} [thb]
\begin{center}
\mbox{\input epsf
 \epsfxsize5.0in \epsfbox{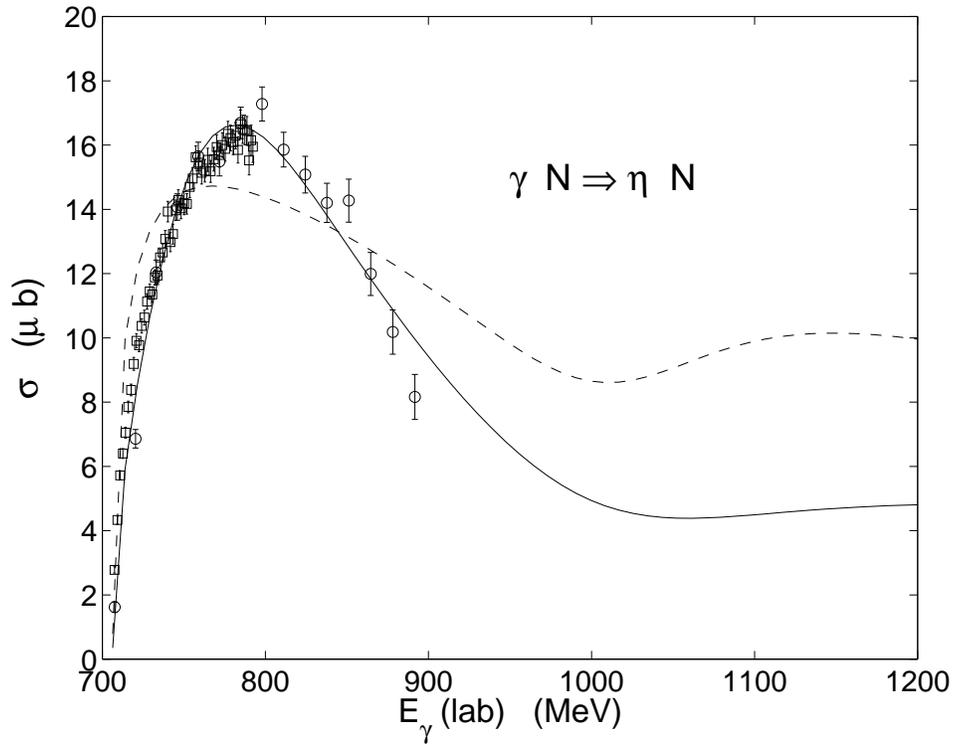}}
 \end{center}
\caption{\label{abbs11photoeta}
  The total eta photoproduction cross section with resonance
  $S_{11}(1535)$ (---) and without $S_{11}(1535)$ (-~-~-). }
\end{figure}

\begin{figure}[thb]
\begin{center}
\mbox{\input epsf
 \epsfxsize5.0in \epsfbox{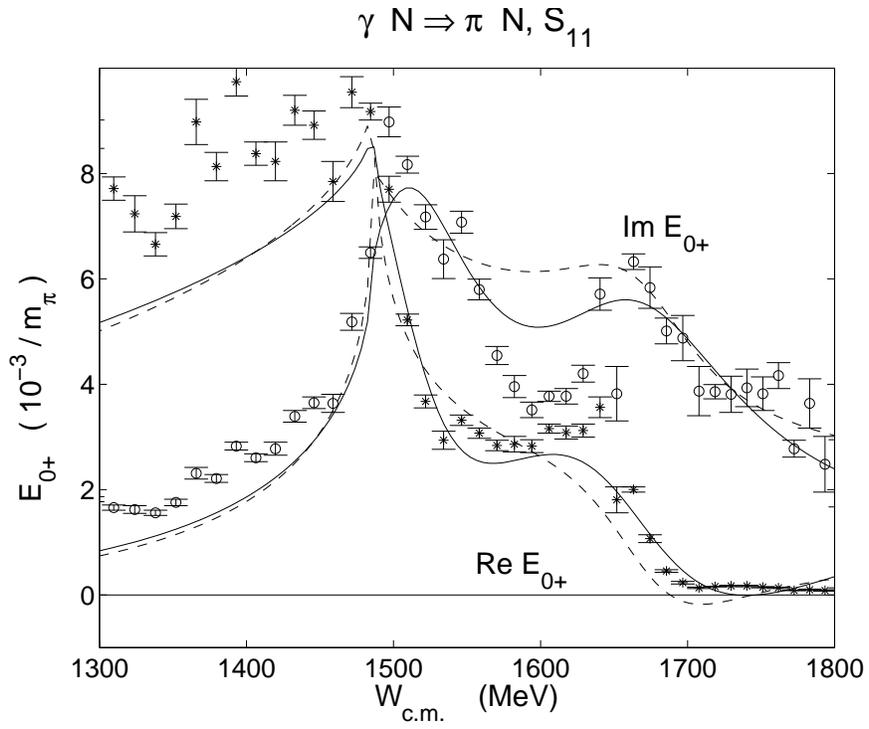}}
 \end{center}
\caption{\label{abbrealims11}
  Real- and imaginary part of the amplitude $E_{0+}$ with resonance
  $S_{11}(1535)$ (---) and without $S_{11}(1535)$ (-~-~-). }
\end{figure}

To first order approximation of perturbation theory, the scattering
matrix $T_{fi}$ is then
\begin{eqnarray}
  T_{fi} &=& _{-}\!\langle N, M_{f}(\vec k) \mid H_{\gamma} \mid N,
  \gamma(\vec \nu), \vec \epsilon \rangle \,,
\end{eqnarray}
and $ _{-}\!\langle N, M_{f} \mid $ is the outgoing scattering state
as calculated in the previous section,
\begin{eqnarray}
  _{-}\!\langle N, M_{f} \mid = \sum_{j} \beta^{f}_{j} \langle R_{j}
  \mid + \sum_{l} \int d^{3}k ~\alpha^{f}_{l}(\vec k) ~\langle N(-\vec
  k),M_{l}(\vec k) \mid .
\end{eqnarray}
The coefficients $\alpha$ and $\beta$ are given in Eqs. (\ref{glbeta})
and (\ref{glalpha}). As before, we consider the two models with and
without the resonance $S_{11}(1535)$, with their respective parameters
and coupling constants given in Tab. \ref{tableconstants} for pion
scattering. We fitted to the eta photoproduction cross section and
pion photoproduction amplitude $E_{0+}$ (see Tab.
\ref{tabpolstelle2}). The experimental data are taken from Wilhelm
\cite{Wil93} and Krusche \cite{Kr95} for eta photoproduction and for
pion photoproduction we used the VPI solution (Sp 95) \cite{ArnRo}.
The results are shown in Figs. \ref{abbs11photoeta} and
\ref{abbrealims11}. The fit without resonance is not as good as the
fit with the resonance. This is partially due to the smaller number of
fitting parameters (five parameters less). However, neither model can
well describe the dip of Im$(E_{0+})$ at 1600 MeV. A more realistic
description of the background would be necessary to better describe
the data below $\eta$ threshold. This would also improve the result
for eta photoproduction for the case without the $S_{11}(1535)$
resonance.

\section{Conclusion}

We have constructed a generalized Lee model for pion scattering and
eta production in the s-wave channel. Using only the Hilbert subspace
corresponding to the quantum numbers of the partial wave $S_{11}$, the
particles that appear in the model are the resonances $S_{11}(1535)$
and $S_{11}(1650)$, the nucleon, the pion and the eta meson. These
five particles interact via a resonant and a nonresonant interaction.
The absence of antiparticles makes it a nonrelativistic model.
However, it has the advantage of being fully unitary and analytically
solvable. Eta and pion photoproduction were additionally calculated
using a perturbation approach.  The coupling constants and other
parameters of the model were determined by fitting the model to
partial wave analyses and cross sections. The experimental data we
used were total cross sections for eta meson production in pion
scattering \cite{Nef92} and eta photoproduction \cite{Wil93}.  In
addition we employed a partial wave analysis of pion nucleon
scattering and $E_{0+} $ amplitudes for pion photoproduction
\cite{ArnRo}, \cite{Kr95}.

We were interested in studying the cusp effect and in obtaining a
better understanding of the interplay of the resonance $S_{11}(1535)$
and the cusp at eta threshold. This is particularly of importance
when determining why the neighboring resonances $S_{11}(1535)$ and
$S_{11}(1650)$, having the same quantum numbers, show quite different
behavior: $S_{11}(1535)$ couples strongly to the eta meson while the
resonance $S_{11}(1650)$ couples to it only very weakly. Following
H\"ohler's statement \cite{Hoh92} that partial wave analyses show no
evident signature for the resonance $S_{11}(1535)$, we investigated a
Lee model with and without the resonance $S_{11}(1535)$.

We found that using the Lee model the experimental scattering data for
various scattering channels could be qualitatively reproduced without
introducing the $S_{11}(1535)$ resonance. Therefore we think that the
resonance $S_{11}(1535)$ is weaker and less important than generally
accepted.

The S-matrix, together with its Riemann sheet structure, was studied.
Eta threshold is the branch point where the four Riemann sheets meet
that were of importance in our discussion. We found a specific
distribution of poles on these sheets that determines the scattering
amplitude on the physical axis. Poles could be attributed to
resonance peak or the cusp peak. More precisely a pole in the
$4^{th}$ sheet that until now has never been taken into account,
determines the shape of the cusp.

When doing calculations close to eta threshold, rescattering and full
calculation to all orders are important.

\acknowledgments

We would like to thank Prof. G. H\"ohler for very fruitful discussions.
One of the authors (J. D.) would like to thank N. E. Hecker for
stimulating discussions and proofreading.  This work was supported by
the Deutsche Forschungsgemeinschaft (SFB201).


\begin{table}
\caption{\protect\label{tableconstants} Coupling constants and masses
of our two models. }
\begin{tabular}{c|cccccccccc}
\multicolumn{11}{c}{\bf Model with resonance $S_{11}$(1535) } \\
\hline
  couplings & $g^{R}_{1,\eta}$ & $g^{R}_{1,\pi}$
 & $g^{R}_{2,\eta}$ & $g^{R}_{2,\pi}$ & $g^{R}_{1,2\pi}$ & $g^{R}_{2,2\pi}$
 & $g_{\eta,\eta}$ & $g_{\eta,\pi}$ & $g_{\pi,\pi}$ & $g_{\pi,2\pi}$ \\ 
 values & 0.24 & 0.12 & 0.01 & 0.25 & 0.0023 & 0.062 &
 2.3$\times10^{-4}$ & 0.01 & 0.0031 & $ 0.024 $ \\ \hline
 \multicolumn{11}{c}{ $ m_{R1} = 1616 \mbox{ MeV},\quad m_{R2} = 1713
   \mbox{ MeV}. $ } \\ 
\end{tabular}
\begin{tabular}{c|cccccccccc}
  \multicolumn{11}{c}{\bf Model without resonance $S_{11}$(1535) } \\ 
  \hline couplings & $g^{R}_{1,\eta}$ & $g^{R}_{1,\pi}$ &
  $g^{R}_{2,\eta}$ & $g^{R}_{2,\pi}$ & $g^{R}_{1,2\pi}$ &
  $g^{R}_{2,2\pi}$
   & $g_{\eta,\eta}$ & $g_{\eta,\pi}$& $g_{\pi,\pi}$  & $g_{\pi,2\pi}$ \\  
   values & 0 & 0 & 0 & -0.24 & 0 & 0.052 & $-0.052 $ & $ 0.045 $ &
   $0.0083$ & $-0.0098 $ \\ \hline
  \multicolumn{11}{c}{ $   m_{R2} = 1704 \mbox{ MeV}. $ } \\ 
\end{tabular}
\end{table}

\begin{table}[bht]
\caption{\protect{\label{tabpolstelle}} Pole positions on the
Riemann sheets for our two models.
 }
\begin{tabular}{l|cc|cc} 
 model & \multicolumn{2}{c|}{$S_{11}(1650)$} &
         \multicolumn{2}{c}{Cusp/$S_{11}(1535)$ } \\
 & 2. sheet & 3. sheet & 3. sheet & 4. sheet \\ \hline
 with $S_{11}(1535)$ \mbox{\hspace{1cm} }& (1655;-111) & (1662;-87) & (1501,-61) & (1528;35)\\ \hline
 no $S_{11}(1535)$ & (1652;-90) & (1670;-60)& - &(1458;57) \\ 
\end{tabular}
\end{table}

\begin{table}[hbt]
\caption{\protect{\label{tabpolstelle2}} Electromagnetic coupling constants for
  eta and pion photoproduction.}
\begin{tabular}{l|cccc} 
 coupling constants \mbox{\hspace{1cm} } &$g^{R}_{1,\gamma}$ & $g^{R}_{2,\gamma}$
  &$g_{\pi, \gamma}$ & $g_{\eta, \gamma}$  \\ \hline
 with $S_{11}(1535)$ & $-2.18 $ & $1.58$ & 0.0027
  & -0.0011 \\ 
 without $S_{11}(1535)$ & $0$ & 1.3  & 0.0024 & -0.0058\\ 
\end{tabular}
\end{table}


\end{document}